\documentclass[twocolumn]{aastex62}
\usepackage{xfrac}
\usepackage{amsmath}

\submitjournal{ApJ}

\shorttitle{Natively Periodic FMM}
\shortauthors{Gnedin}

\def\vec#1{\mathbf{#1}}
\def\dvec#1{{\rm d}\vec{#1}}

\def\oct#1{\sfrac{#1}{8}}

\def\ifft{\underset{k\rightarrow x}{\widehat{\bf F}}}

\begin{document}
\title{Natively Periodic Fast Multipole Method: Approximating the Optimal Green's Function}

\correspondingauthor{Nickolay Y.\ Gnedin}
\email{gnedin@fnal.gov}

\author{Nickolay Y.\ Gnedin}
\affiliation{Fermi National Accelerator Laboratory;
Batavia, IL 60510, USA}
\affiliation{Kavli Institute for Cosmological Physics;
The University of Chicago;
Chicago, IL 60637 USA}
\affiliation{Department of Astronomy \& Astrophysics; 
The University of Chicago; 
Chicago, IL 60637 USA}

\begin{abstract}
The Fast Multipole Method (FMM) obeys periodic boundary conditions "natively" if it uses a periodic Green's function for computing the multipole expansion in the interaction zone of each FMM oct-tree node. One can define the "optimal" Green's function for such a method that results in the numerical solution that converges to the equivalent Particle-Mesh solution in the limit of sufficiently high order of multipoles. A discrete functional equation for the optimal Green's function can be derived, but is not practically useful as methods for its solution are not known. Instead, this paper presents an approximation for the optimal Green's function that is accurate to better than $10^{-3}$ in $L_{\rm MAX}$ norm and $10^{-4}$ in $L_2$ norm for practically useful multipole counts. Such an approximately optimal Green's function offers a practical way for implementing FMM with periodic boundary conditions "natively", without the need to compute lattice sums or to rely on hybrid FMM-PM approaches.
\end{abstract}


\section{Introduction}
\label{sec:intro}

Numerical simulations most commonly model a limited spatial volume, but actual space has no limits; hence, in many astrophysical applications numerical simulations have to employ periodic boundary conditions as a makeshift representation of unlimited space. Given an approximate representation of physical reality (discussing pros and cons of such an approximation is not a subject of this paper), one can then address a mathematically exact question of implementing periodic boundary conditions in a computational algorithm. For gravity calculations, imposing periodicity is not necessarily trivial. It is, of course, trivial for the simple Particle-Mesh (PM) method \citep{he1988} that uses the discrete Fast Fourier Transform (FFT) to compute the gravitational potential (and/or accelerations) on the uniform grid. The discreteness of the FFT ensures that scales larger than the computational box size (i.e.\ waves with spatial frequencies below the fundamental frequency) do not contribute to the Green's function $G$ that is used to compute the potential from the assigned density on the grid.

In modern simulations multipole-based methods are commonly used to compute the gravitational accelerations for particles or for cells on a grid, such as a classical Barnes-Hut tree method  \citep{bh1986} or a Fast Multipole Method (FMM) introduced by Greengard \& Rokhlin \citep{gr1987,gr1997,cgr1999}. In cosmological simulations the problem of imposing periodic boundary conditions for Tree or FMM methods has been traditionally solved by using hybrid Tree-PM and FMM-PM approaches, as implemented in, for example, the widely used cosmological code GADGET \citep{gadget2,gadget4}. In such methods the gravitational force is split into a short-range part that is solved by a Tree or FMM and a long-range part that is solved on a uniform grid with a PM method. Because the two parts of the full gravitational force are solved with two different methods, an error is always introduced at scales where the two components are comparable. For example, in GADGET such an error can reach $\sim$1\% \citep{gadget2,gadget4}. It is not presently known if that error results in simulation artifacts.

Since FMM is computing a convolution of the density distribution with a Green's function, one can simply use a periodic Green's function to compute the convolution with periodic boundary conditions, in a direct analogy to PM. Such an FMM method would have periodic boundary conditions "natively", without any additional computations. This is, of course, not a new idea and has been proposed in the past \citep[c.f.][]{ys2018}. A choice, however, needs to be made of which Green's function to use.

While in the continuous limit in 3D there is just one Green's function for the gravitational potential,
\[
    G_0(\vec{x}) = -\frac{1}{4\pi r}
\]
(or $\tilde{G}_0(\vec{k}) = -1/k^2$ in Fourier space - hereafter I use a tilde symbol to label a Fourier transform of a function), this is not so for a discrete problem. In fact, on a uniform grid there are infinitely many Green's functions that approach $G_0$ in the continuous limit. One such function is the "exact-in-Fourier-space" Green's function,
\begin{equation}
    \tilde{G}_K(\vec{k}) = \begin{cases}
    -1/k^2, & \vec{k}\neq 0 \\
    0, & \vec{k} = 0
    \end{cases},
    \label{eq:gfk}
\end{equation}
\[
    G_K(\vec{x}) = \hat{\bf F}_{k\rightarrow x}\left[\tilde{G}_K(\vec{k})\right],
\]
where the operator $\ifft[...]$ denotes the (inverse) Fourier transform of the argument in square brackets. Now $\vec{k}$ takes only discrete values, $\vec{k}=(2\pi/L)\vec{n}$, where $\vec{n}$ is the integer-valued vector on an $N^3$ grid, $\vec{n}=(n_x,n_y,n_z)$ with $n_i=-N/2+1,-N/2+2,...,-1,0,1,...,N/2-1,N/2$. Another extreme is the "exact-in-real-space" Green's function,
\begin{equation}
    G_X(\vec{x}) = -\frac{1}{4\pi} g\left(\left|\vec{x}\diamond L\right|\right),
    \label{eq:gfx}
\end{equation}
where $g(r)=1/r$ in 3D and $g(r)=\log(r^2)$ in 2D and the symbol $\diamond$ denotes the periodic coordinate wrap in a box of size $L$,
\[
    x \diamond L \equiv x - L\, {\rm nint}(x/L),
\]
where nint() is the function returning the nearest integer to a real number. A particular feature of $G_X$ is that it is continuous but not differentiable at $|x|=L/2$.

\begin{figure}[t]
\includegraphics[width=\hsize]{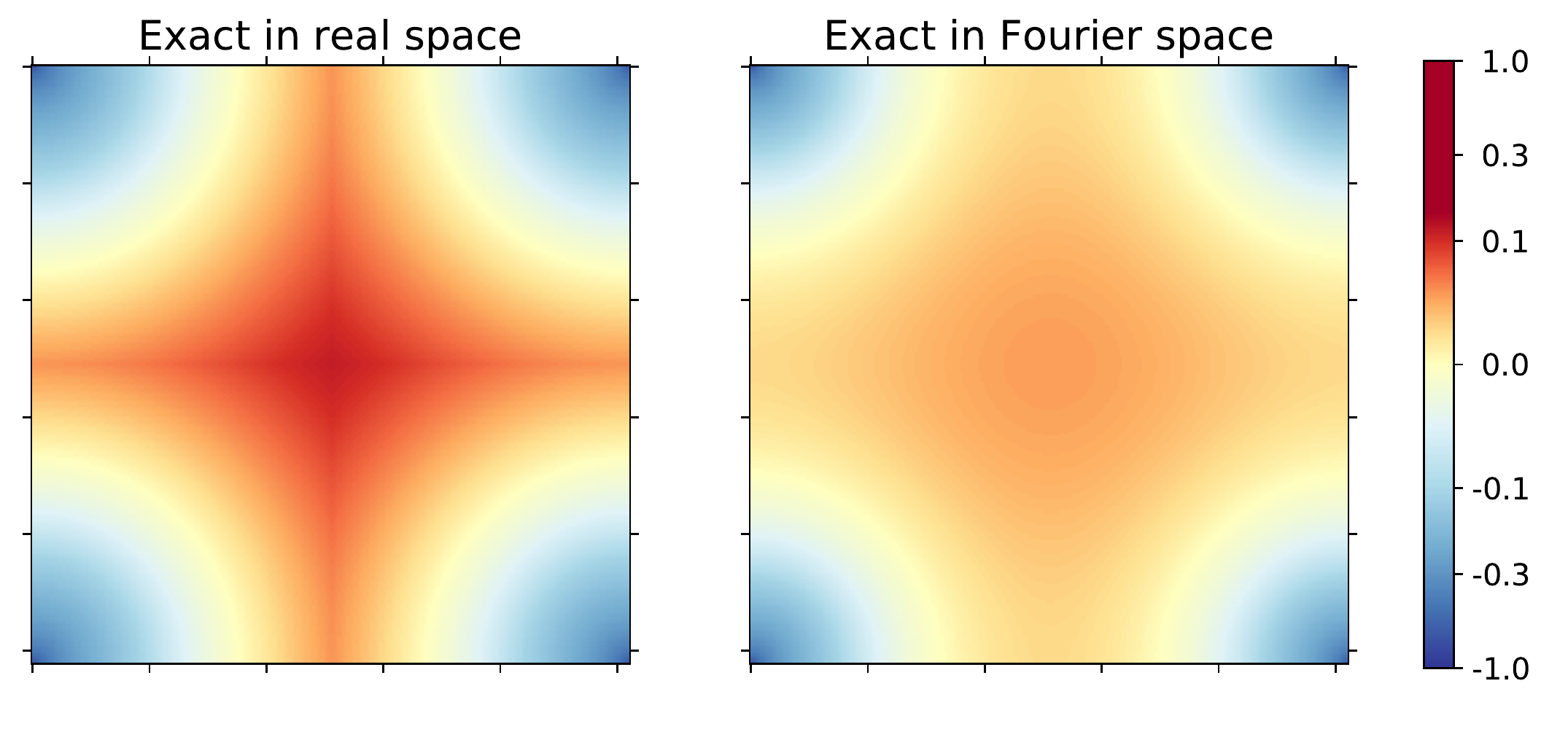}
\caption{Examples of the "exact-in-real-" and the "exact-in-Fourier-space" Green's functions in 2D in real space. \label{fig:gfs} }
\end{figure}

These two examples of Green's functions for a 2D case are shown in Figure \ref{fig:gfs}. One can construct infinitely many Green's functions between these two extremes. For example, the Green's function that enters the commonly used Ewald summation technique \citep{e1921} is simply 
\begin{eqnarray}
  G_E(\vec{x}) & = & G_X(\vec{x}) {\rm erfc}(r/r_S) + \frac{1}{2\pi^{3/2}r_s} + \nonumber\\
  & & \ifft\left[\tilde{G}_K(\vec{k}) e^{-k^2r_S^2/4}\right]  \nonumber 
\end{eqnarray}
for some $r_S \ll L$ (in which case additional terms due to periodic images are negligible). The Ewald summation Green's function approaches $G_X$ (up to a constant) for $r\ll r_S$ and $G_K$ for $r\gg r_S$.

\section{Natively Periodic FMM}

The FMM algorithm uses an oct-tree (in 3D) to tessellate the space. In what follows I assume that the tree has at least 3 levels of refinement as the general case - special cases of shallower trees can be considered similarly. For levels above 3 no assumptions about the structure of the oct-tree need to be made, the tree structure can be completely arbitrary as a generic FMM algorithm does not need to make any assumption about spatial refinement \citep[c.f.][]{Hrycak1998}.

\begin{figure}[t]
\includegraphics[width=\hsize]{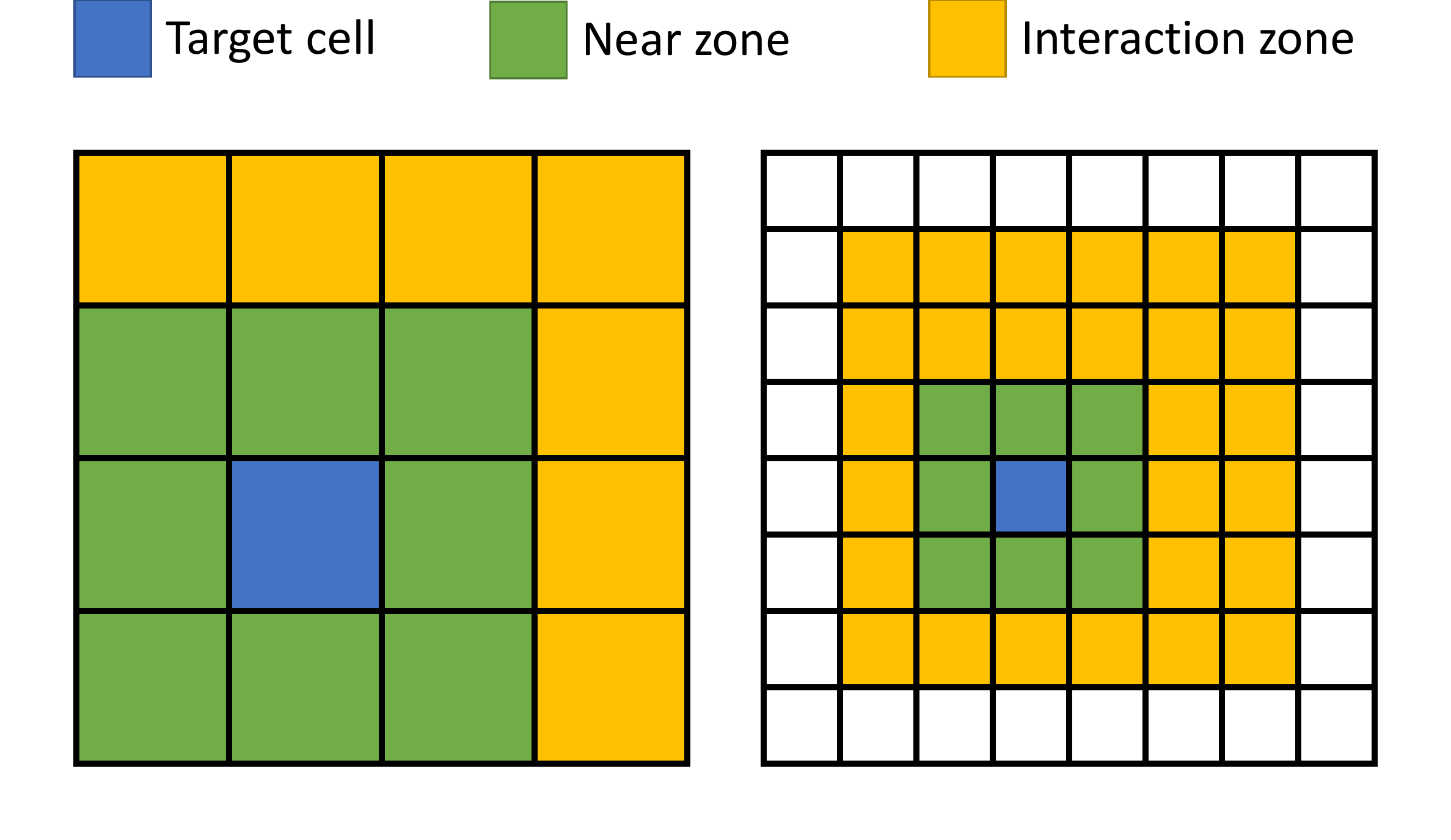}
\caption{Levels 2 and 3 of the FMM oct-tree. The immediate neighbors (green) of the target node (blue) form its "near zone", while the rest of the children-of-its-parent-neighbors (orange nodes) form the "interaction zone". On level 3 the interaction zone is entirely within the computational domain, and hence is oblivious to the nature of boundary conditions. Note, that with periodic boundary conditions the target (blue) nodes can always be shifted to the locations shown in the figure by a translation followed by a periodic wrap (i.e.\ this figure shows a generic case and no special cases, like a corner node, exist under periodic boundary conditions).\label{fig:fmm} }
\end{figure}

Levels 2 and 3 of such a tree are shown in Figure \ref{fig:fmm}. Gravitational potential for all resolution elements (particles or mesh cells) in the target oct-tree node (blue square) is computed as a sum of 3 parts: (1) the contribution from the "interaction zone" (all children of the parent neighbors that are not neighboring the target node) is computed from the multipole expansion in those nodes; (2) the contribution from the parent multipoles, which, in turn, is a sum of parent interaction zone and the grand-parent multipole contribution and which, in its turn, is a sum of grant-parent interaction zone and the grand-grand-parent multipole contribution, etc, and (3) the contribution from the "near zone" (immediate neighbors of the target node), which either is added exactly for a leaf node by, for example, a direct summation or a PM solver with non-periodic boundary conditions or is resolved further unto children, grand-children etc interaction + near zones for non-leaf nodes.

On all levels 3 and above the interaction zone is entirely within the computational domain (after an appropriate periodic translation), so as long as the neighbor selection accounts for the periodic boundary conditions, it is possible to use the original, non-periodic FMM algorithm. The effect of periodic boundary conditions only needs to be accounted for explicitly at level 2, and after that it is passed to the rest of the tree through the parent-to-child multipole transformation automatically.

Hereafter, the particular implementation of the FMM algorithm that is going to be used is a "Hierarchical Particle-Mesh" (HPM) approach introduced in \citet{g2019}. The HPM flavor of the FMM algorithm generally follows the \citet{va2010} implementation of FMM with Cartesian multipoles and replaces the computation of the interaction zone contribution with a single FFT convolution. In order to do that, it replaces Cartesian multipoles with a uniform mini-grid (a "gridlet") of effective masses so that the Cartesian multipoles of the gridlet mass distribution match the original multipoles exactly up to the given order $N_g$. Following the widely used "multi-index" notation in which for 3D vectors $\vec{n}=(n_x,n_y,n_z)$ and $\vec{r}=(x,y,z)$
\[
    \begin{array}{ll}
    \vec{n}! & \equiv n_x!\,n_y!\,n_z!, \\
    \vec{r}^\vec{n} & \equiv x^{n_x}\,y^{n_y}\,z^{n_z}, \mathrm{and}\\
    \dvec{r} & \equiv dx\,dy\,dz,
    \end{array}
\]
the Cartesian multipoles of a mass distribution $\rho(\vec{r})$ in one FMM node can be expressed as
\[
    Q_\vec{n} \equiv \frac{1}{\vec{n}!} \int \rho(\vec{r}) \vec{r}^\vec{n} \dvec{r}
\]
and the corresponding effective masses $M_\vec{p}$ are then solutions to the following linear equation:
\[
    Q_\vec{n} = \frac{1}{\vec{n}!} \sum_{\vec{p}=0}^{N_g-1} M_\vec{p} \vec{r}_\vec{p}^\vec{n},
\]
where $\vec{r}_\vec{p}$ are coordinates of the uniform mini-mesh covering the FMM node and $\vec{r}_\vec{p}^\vec{n}$ is a Vandermonde matrix.

Effective masses from all nodes in the interaction zone are placed on a common uniform grid of size $(6N_g)^D$ ($D$ is the number of spatial dimensions) and the values of the gravitational potential in the target node are computed with a single FFT similar to the PM method. Hence, the only change needed to make HPM natively periodic is to use the appropriate Green's function for the FFT on the interaction zone at level 2 (yellow nodes in Fig.\ \ref{fig:fmm}).

\section{Optimal Green's Function}

\begin{figure*}[t]
\includegraphics[width=\hsize]{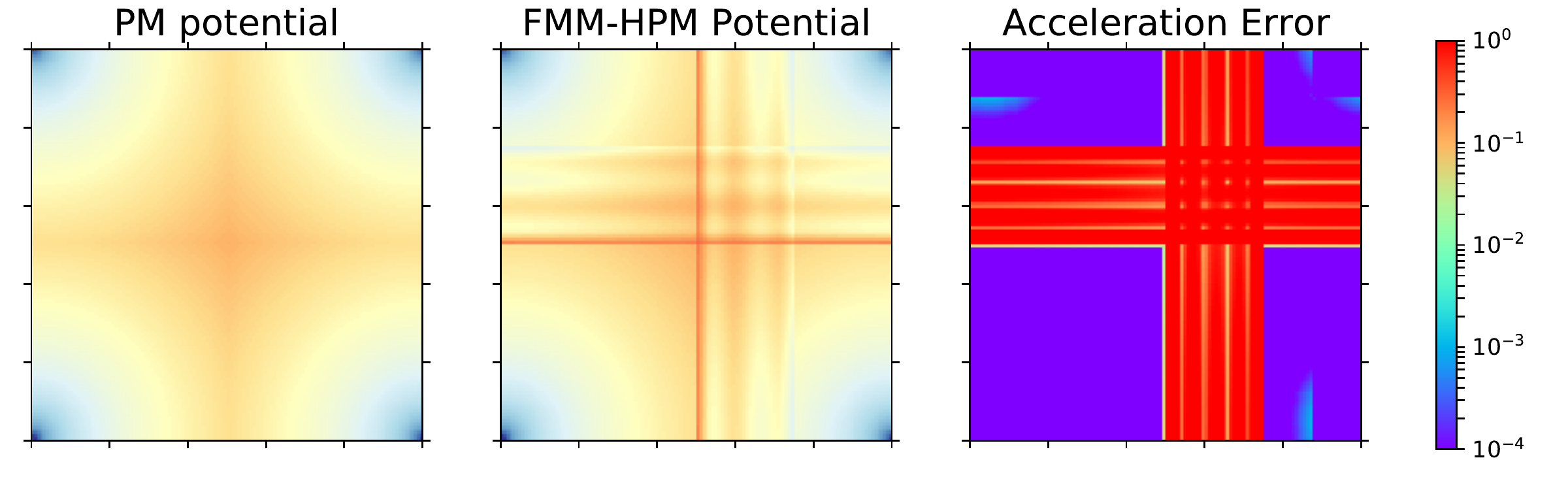}
\caption{Solutions in 2D for the gravitational potential from a point source located in the lower-left corner of the box with the "exact-in-real-space" Green's function (Eq.\ \ref{eq:gfx}) and $N_g=6$ gridlet ($6^3$ cartesian multipoles). The left panel shows the PM method, while the middle panel shows the FMM-HPM method. The right panel shows the difference in the gravitational acceleration between the two, which in the worst case exceeds 100\% due to the catastrophic aliasing in the "exact-in-real-space" Green's function $G_X$.\label{fig:errex}}
\end{figure*}

The choice of the "appropriate Green's function" is not trivial, however. As an example, Figure \ref{fig:errex} shows solutions for the gravitational potential by the PM and FMM-HPM methods with the "exact-in-real-space" Green's function $G_X$ (Eq.\ \ref{eq:gfx}). For this test the FMM-HPM method is implemented with the code described in \citet{g2019}. The oct-tree is at least 3 levels deep (with just 2 levels the PM and FMM-HPM results are identical by construction). Each FMM oct-tree node is covered by an $16^2$ sub-grid (for this 2D test, and by $16^3$ sub-grid for the 3D tests presented below) - the choice for the size of sub-grid is dictated by the requirement that it is significantly larger than the largest number of multipoles used in tests below; otherwise there is an accidental cancellation of errors that misrepresents the real accuracy of FMM. Thus, the total box is covered by the uniform $128^D$ grid ($D$ is the dimension of space), which is the grid used by the PM method. The gravitational source is modeled by assigning the density of $1/128^D$ in the most lower-left cell of the uniform grid, so that the total mass of the source is 1 if the box size is taken to be 1. A number of other, randomly chosen locations were tested, and the corner source was found to result in the largest error possible (i.e.\ it is the worst case scenario - this is also illustrated below). 

The definition of the error between PM and FMM-HPM also requires special consideration. In the PM method gravitational accelerations are computed from the potential by the 4th order finite difference (up to the 8th order was tested, the 4th order was found to give a converged answer), while in the FMM the accelerations are computed from the multipole expansion. The difference between the accelerations from FMM and from PM is made of these 2 contributions: (a) the error due to the choice of the Green's function and (b) the error in the multipole expansion. In order to only compute (a), the error in the multipole expansion is estimated by comparing FMM and PM in the non-periodic case. Hence, the error due to the choice of the Green's function is computed as
\begin{equation}
    \Delta g = | \vec{g}_{\rm FMM}^{\rm P} - \vec{g}_{\rm PM}^{\rm P} - (\vec{g}_{\rm FMM}^{\rm NP} - \vec{g}_{\rm PM}^{\rm NP}) |,
    \label{eq:gerr}
\end{equation}
where superscripts P and NP stand for "periodic" and "non-periodic". The error due to the multipole expansion is also shown below in Fig.\ \ref{fig:eng}. Such a correction is not perfect, of course, hence the errors due to the choice of the Green's function presented below may be somewhat overestimated.

Ideally, the relative error would be most informative. However, for a single source in a box with periodic boundary conditions there is a location where the gravitational acceleration vanishes identically and planes where each of the components of the acceleration vanishes. Hence, small absolute errors near such locations result in large relative errors. The search for the optimal Green's function below utilizes minimization of some norm of the error. In order to avoid the error to be dominated by just one location in the computational box, the absolute error is used hereafter. In a unit box with $GM=1$ the characteristic gravitational acceleration is 1, so the absolute error measures the relative error with respect to that characteristic acceleration. 

The absolute error between the PM and FMM-HPM when both use the "exact-in-real-space" Green's functions $G_X$ is shown in the right panel of Fig.\ \ref{fig:errex}. That error is very large, primarily due to sharp features in the gravitational potential from the FMM-HPM. These features arise from aliasing in the "exact-in-real-space" Green's function $G_X$, since, as has been mentioned above, $G_X$ is not differentiable at the distance of half the box size.

While the two solutions are so different, they both are valid solutions of the Poisson equation with periodic boundary conditions. Intuitively, however, it seems that the PM solution is in some sense "better". While this is not a mathematically rigorous statement, I am going to adopt it as an ansatz in the rest of this paper. One can then define an "optimal" Green's function $G_*$ for the natively periodic FMM as \emph{the Green's function with which the PM method and FMM produce the same gravitational potential}. 

A reasonable question to ask is why the two Green's functions should be the same. For example, one can use a PM method with the "exact-in-Fourier-space" Green's function $G_K$ and then choose the Green's function for FMM such that the solutions agree. That would not be possible, however. For the solutions to match, the FMM solution in the near zone should match too. For a regular grid, this can be solved with a PM-like method, using FFT with non-periodic boundary conditions (which are most easily achieved by doubling the grid size - this is the approach taken in this paper). If the Green's function used in the global PM solution does not match exactly the one used in the near zone part of the FMM computation, the two solutions would differ near the source, precisely where the force is the largest, resulting in a large absolute error even from small relative errors.

The requirement that the PM and the FMM solutions agree is not mathematically rigorous due to the existence of additional errors due to multipole expansion - a realistic realization of FMM would never agree with the PM solution exactly because the FMM solution is (almost) always approximate. One can derive a mathematically rigorous result in the limit when the FMM uses Cartesian multipoles of the order equal to the size of the sub-grid. In that special case there is no error due to the multipole expansion in the interaction zone.

In order to derive the equation for the "optimal" Green's function $G_*$, let us consider the 3 level deep oct-tree. Given the distribution of masses $M_\vec{p}$ on the gridlet in a level 3 node, the PM solution for the potential in some other oct-tree node is simply a convolution over that node gridlet (because in the special case considered the size of the gridlet is equal to the size of the sub-grid in that oct-tree node),
\begin{equation}
    \Phi_\vec{p}^{\rm PM} = \sum_{\vec{q}} G_*\left(\vec{d}+\Delta x(\vec{p}-\vec{q})\right) M_\vec{q},
    \label{eq:pmconv}
\end{equation}
where $\vec{p}$ and $\vec{q}$ are vector-valued indices over the gridlets, $\vec{p}=(p_x,p_y,p_z)$, $p_i=0,...,N_g-1$, with $N_g$ being the gridlet size (6 in the test shown in Fig.\ \ref{fig:errex}); $\Delta x=1/(8 N_g)$ is the gridlet spacing (the size of one cell on the PM mesh), and $\vec{d}$ is the vector from the center of the source oct-tree node to the target oct-tree node.

The same computation in the FMM-HPM method would take 3 steps:
\begin{enumerate}
    \item projecting the source multipoles from a child node $C_S$ at level 3 ($M_\vec{q}^{L=3}$) to level 2 ($M_\vec{q}^{L=2}$),
    \begin{equation}
        M_\vec{s}^{L=2} = \sum_{\vec{q}} T_{\vec{s}\vec{q}}^{C_S} M_\vec{q}^{L=3},
        \label{eq:fmm1}
    \end{equation}
    where matrices $T_{\vec{s}\vec{q}}$ are defined in \S 2.3 of \citet{g2019};
    \item computing the interaction zone contribution to the multipoles on level2,
     \begin{equation}
        \Phi_\vec{r}^{L=2} = \sum_{\vec{s}} G_*\left(\vec{d}_{L=2}+\Delta x_{L=2}(\vec{r}-\vec{s})\right) M_\vec{s}^{L=2},
        \label{eq:fmm2}
    \end{equation}
    and
    \item projecting the level 2 multipoles up to a child node $C_T$ at level 3,
    \begin{equation}
        \Phi_\vec{p}^{L=3} = \sum_{\vec{r}} T_{\vec{r}\vec{p}}^{C_T} \Phi_\vec{r}^{L=2}.
        \label{eq:fmm3}
    \end{equation}
\end{enumerate}
Comparing Equation (\ref{eq:pmconv}) with Equations (\ref{eq:fmm1}-\ref{eq:fmm3}) and requiring that
$\Phi_\vec{p}^{\rm PM} = \Phi_\vec{p}^{L=3} + {\rm const}$ for any $M_{\vec{q}}$, one arrives at the equation on $G_*$:
\[
    G_*\left(\vec{d}_{L=3}+\Delta x_{L=3}(\vec{p}-\vec{q})\right) = 
\]
\[
    {\rm const} + \sum_{\vec{r},\vec{s}} T_{\vec{r}\vec{p}}^{C_T} T_{\vec{s}\vec{q}}^{C_S} G_*\left(\vec{d}_{L=2}+\Delta x_{L=2}(\vec{r}-\vec{s})\right).
\]
Since $\Delta x_{L=2} = 2 \Delta x_{L=3}$ for an oct-tree and the source and target child nodes are offset from their respective parent node centers by vectors $\vec{a}^{C_S}$ and $\vec{a}^{C_T}$, one can rewrite the equation for $G_*$ as
\[
    G_*\left(\vec{d}+\vec{a}^{C_T}-\vec{a}^{C_S}+\Delta x(\vec{p}-\vec{q})\right) = 
\]
\begin{equation}
    {\rm const} + \sum_{\vec{r},\vec{s}} T_{\vec{r}\vec{p}}^{C_T} T_{\vec{s}\vec{q}}^{C_S} G_*\left(\vec{d}+2\Delta x(\vec{r}-\vec{s})\right)
    \label{eq:opt}
\end{equation}
with $\Delta x$ being the size of the cell on the PM mesh. Notice, that $G_*$ may be a function of the gridlet size $N_g$ (or, equivalently, the number of multipoles $N_g^3$) since for different gridlet sizes Equation (\ref{eq:opt}) is different.

The challenge of solving Equation (\ref{eq:opt}) is two fold: first, it is a discrete functional equation and does not belong to any class of mathematical equations for which methods of solution are known. Second, the vector $\vec{d}$ is the vector from one oct-tree node center at level 2 to another oct-tree node in the first node interaction zone, i.e.\ Equation (\ref{eq:opt}) does not constraint $G_*(\vec{x})$ everywhere in space but only for $\vec{x}$ such that at least one of its components is greater than $L/4$ by absolute value (the later is not a serious limitation if $G_*$ is assumed to be analytic).  

\begin{figure*}[t]
\includegraphics[width=\hsize]{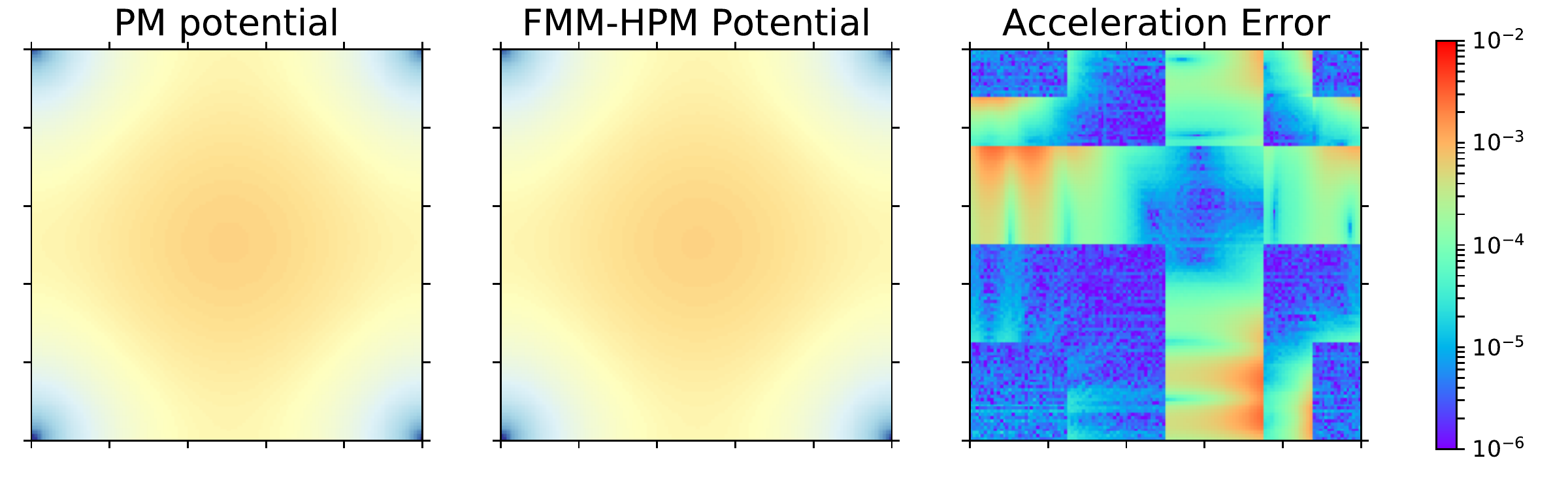}
\caption{Same as Figure \ref{fig:errex}, but for the approximately-optimal Green's function $G_S$. Notice that the scale of the colorbar is reduced 100-fold compared to Figure  \ref{fig:errex}.\label{fig:erropt}}
\end{figure*}

In trying to solve Equation (\ref{eq:opt}) I attempted to use several simple iteration schemes, but found none that converged. An alternative approach to finding an \emph{approximation} to the optimal Green's function is to assume a parameterized functional form for $G_*$ and fit for the best parameter values.

In the following, I adopt the following ansatz:
\[
    G_*(\vec{x}) \approx G_S(\vec{x}) \equiv -\frac{1}{4\pi} g\left(\left|\vec{\hat{x}}\diamond L\right|\right),
\]
where $\hat{x}$ should approach $x$ at small scales to recover the Newtonian force and should be differentiable at half the box size to avoid catastrophic aliasing from Fig.\ \ref{fig:errex}. One such choice for $\hat{x}$ is
\[
    \hat{x} = L \xi_n\left(\frac{x}{L}\right)
\]
with
\[
   \xi_n(t) = \frac{\sin(\pi t)}{\pi} + \sum_{k=1}^{n} a_k\left(\frac{\sin((2k+1)\pi t)}{(2k+1)\pi}-\frac{\sin(\pi t)}{\pi}\right).
\]
The coefficients $a_k$ are free parameters that are found by minimizing the $L_2$ norm of the error between the PM and FMM-HPM solutions (such as the error shown on the right panel of Fig.\ \ref{fig:errex}). The minimization is performed using the \emph{Minuit2} library\footnote{\url{https://root.cern.ch/doc/master/Minuit2Page.html}}. One can also minimize the $L_{\rm MAX}$ norm using a simpler minimization method that does not rely on the existence of the Hessian, such as the downhill simplex method \citep{nm1965} with insignificant change in the best parameter values and the precision of the approximation.

In all cases, however, terms beyond $k=1$ do not give a significant improvement in the precision and almost perfectly correlate with the $k=1$ term; in 3D with large $N_g$ there is a noticeable improvement in $L_{\rm MAX}$ values (Table \ref{tab:ak}), but then the precision of the approximation is already so high that the gain in precision is not worthy the added numerical complexity. Why this is so is clear from Figure \ref{fig:erropt}, that shows the error in the acceleration for the "approximately optimal" Green's function $G_S$ with $n=1$ and the value of the coefficient $a_1$ (as well as the accuracy of the fit) given in Table \ref{tab:ak}. The error is reduced by almost 3 orders of magnitude compared to the "exact-in-real-space" Green's function, but the residual error has a complex structure and is not well approximated by simple analytical functions.

In fact, for all cases considered the value of the first coefficient $a_k$ is sufficiently close to the value $-0.125$ that gives a cancellation of the third order in $x$ term,
\[
    \frac{9}{8}\sin(t) - \frac{1}{8}\frac{\sin(3t)}{3} = t + O(t^5)
\]
for $t\rightarrow0$, so that a single function $G_S$ with $n=1$ and $a_1=-\oct{1}$, (which I will call $G_{S1}$ hereafter) can be used for all dimensions and for all gridlet sizes without significant loss of precision. It is not clear if this is just a coincidence, though, since adding the next cancelling term,
\[
\frac{75}{64}\sin(t) - \frac{25}{128}\frac{\sin(3t)}{3} + \frac{3}{128}\frac{\sin(5t)}{5} = t + O(t^7),
\]
significantly deteriorates the precision. 

\begin{figure*}[t]
\includegraphics[width=0.5\hsize]{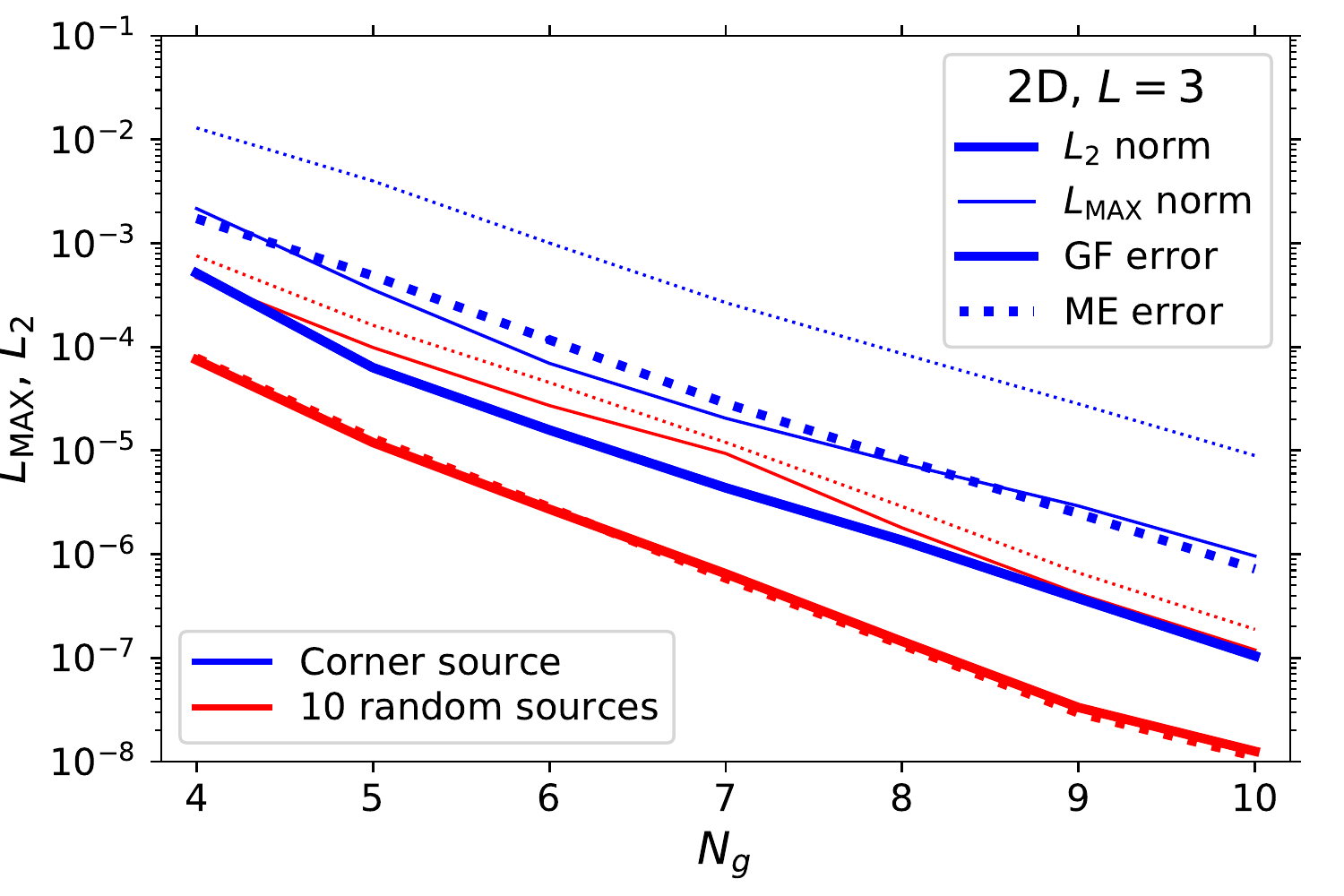}\includegraphics[width=0.5\hsize]{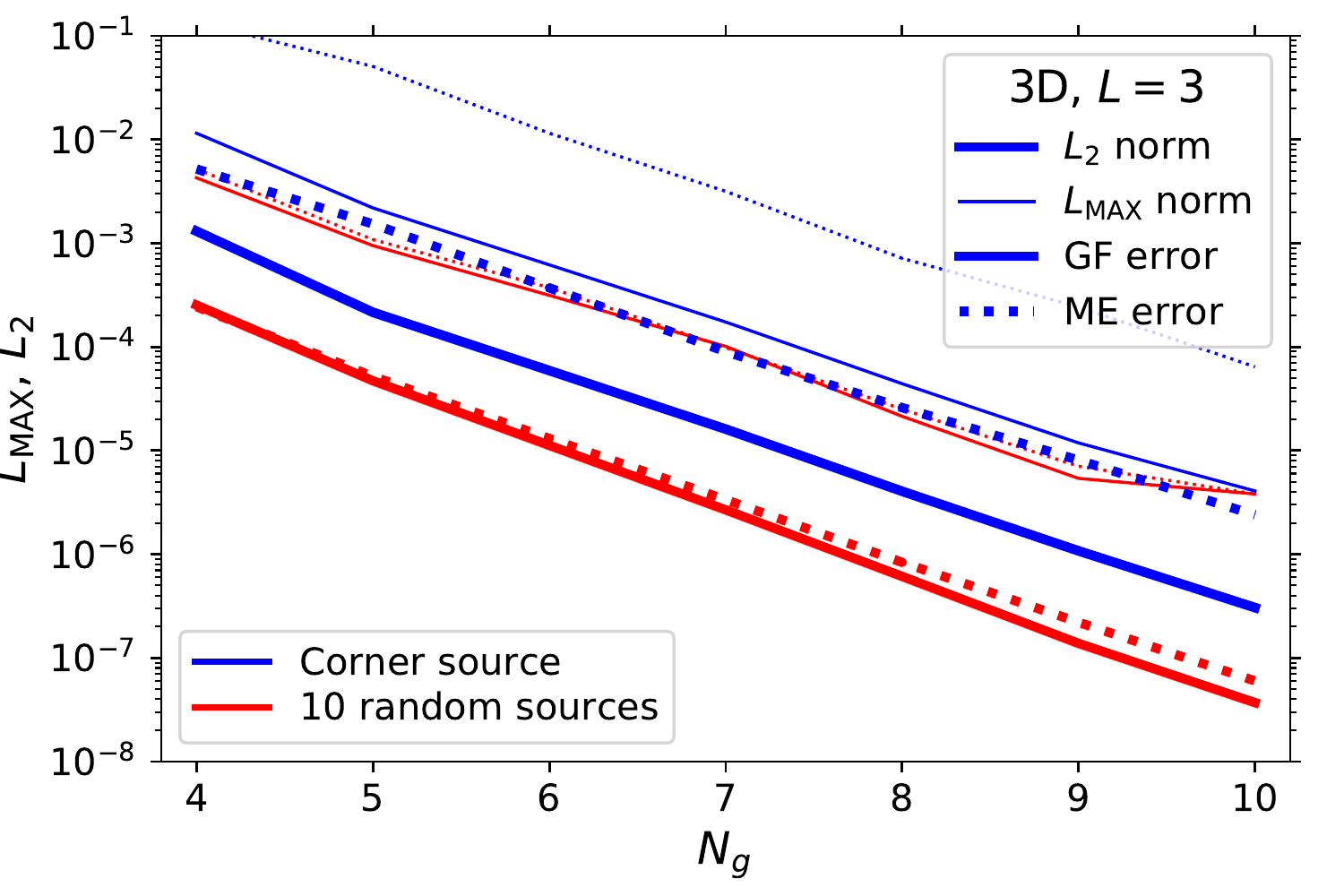}\newline
\includegraphics[width=0.5\hsize]{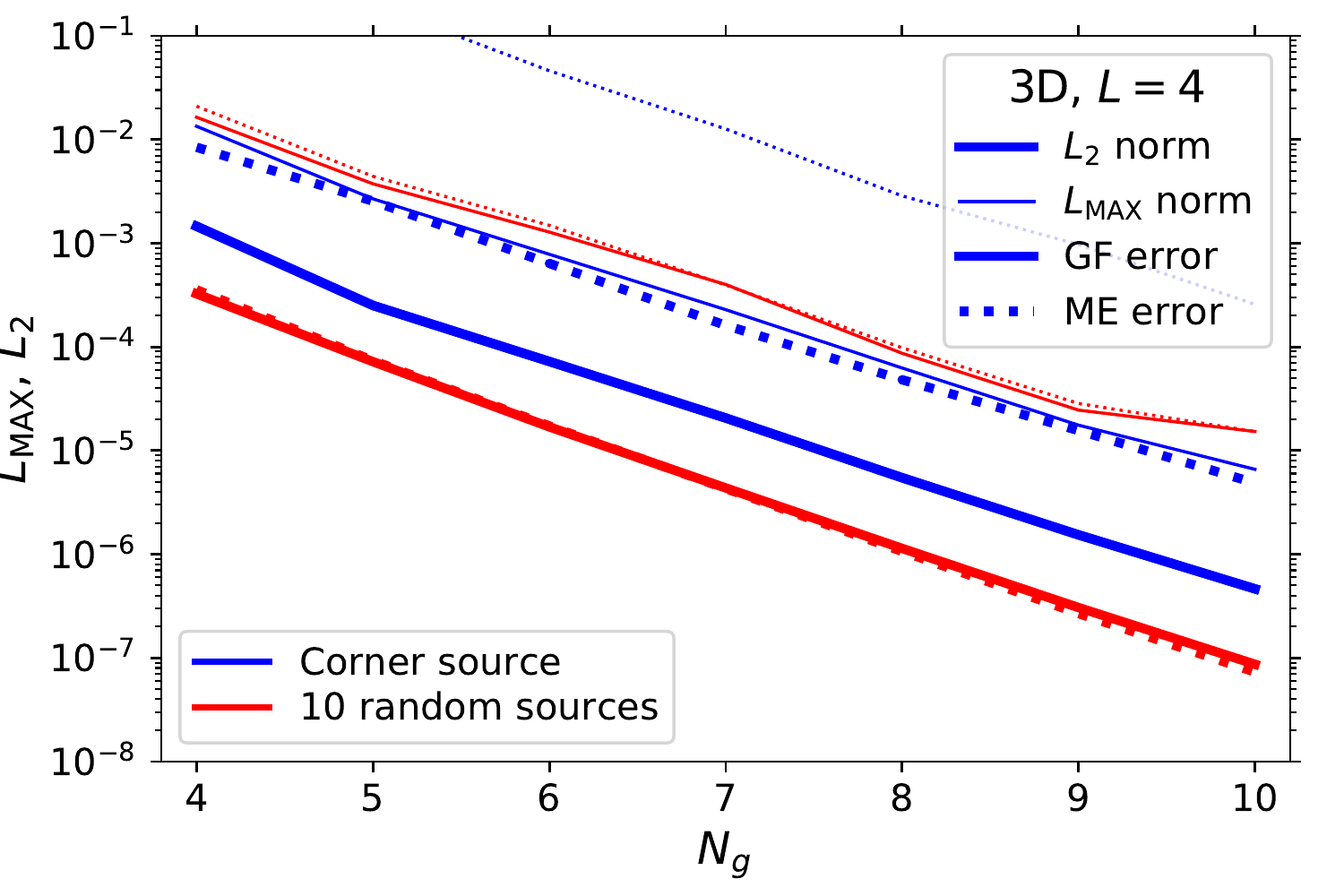}\includegraphics[width=0.5\hsize]{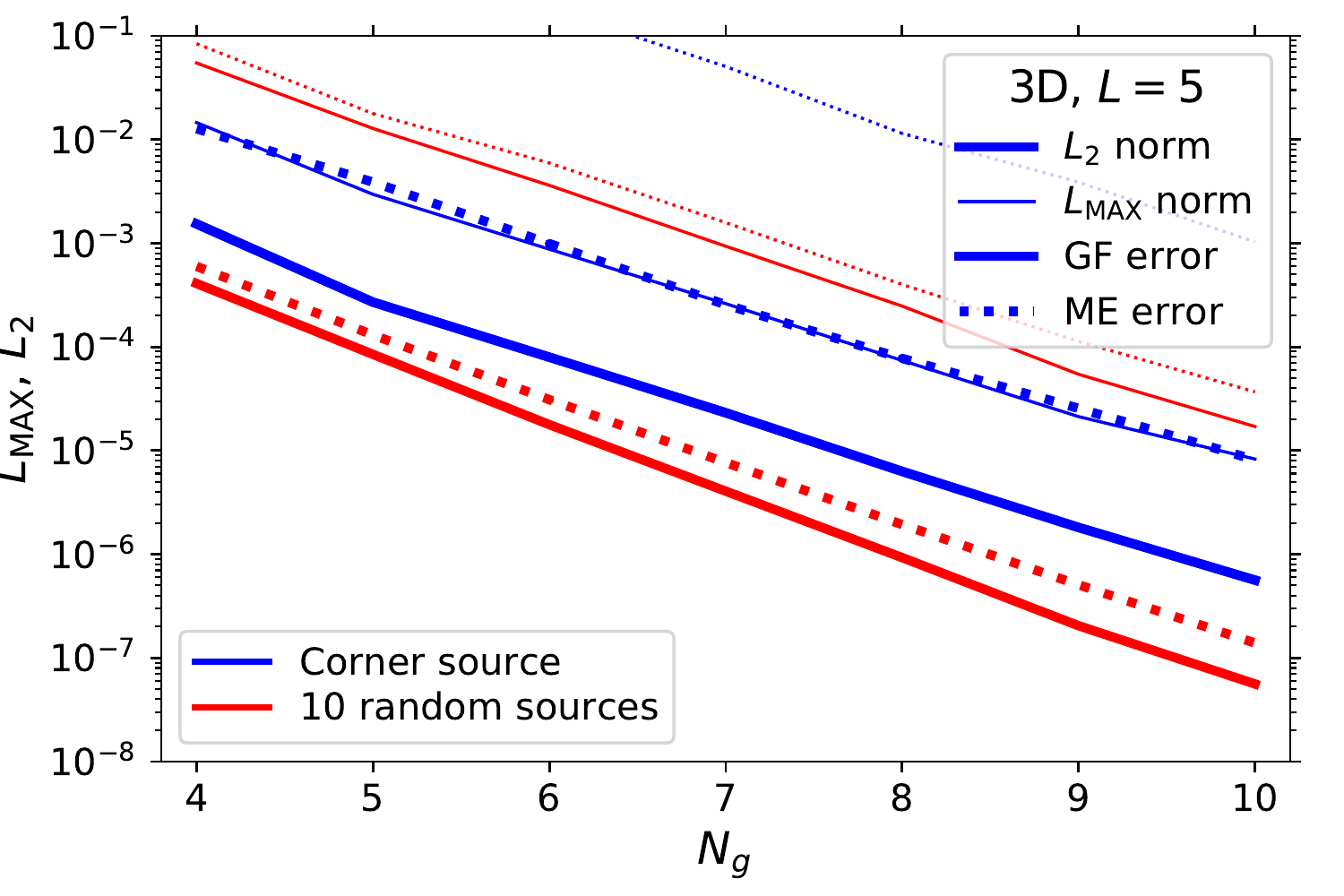}
\caption{$L_2$ and $L_{\rm MAX}$ norms for the approximate Green's function $G_{S1}$ (the "GF error") and for the multipole expansion in non-periodic case (the "ME error") as a function of the gridlet size $N_g$ (the number of Cartesian multipoles is $N_g^3$) for two tests: a single source placed in the corner of the computational box (as in Figs.\ \ref{fig:errex} and \ref{fig:erropt}) and 10 sources placed randomly in the box. The top row shows 2D and 3D tests with 3-level deep FMM tree ($128^D$ PM mesh), the bottom row shows 3D tests for 4- and 5-level FMM trees ($256^3$ and $512^3$ PM meshes). \label{fig:eng}}
\end{figure*}

The performance of the approximately optimal Green's function $G_{S1}$ is shown in Figure \ref{fig:eng} for a number of gridlet sizes and for both 2D and 3D tests. In addition to the worst case scenario of the single corner source, I also show a more realistic case of 10 sources randomly distributed in the computational box, for which the errors due to both approximation for the Green's function (the "GF error") and due to the multipole expansion (the "ME error") are significantly smaller than the worst case. In all tests considered the Green's function error is below or about the error due to the multipole expansion for $N_g\leq10$. For the 10 random source test the error due to the approximate Green's function is similar to the error due to the multipole expansion, indicating that the subtraction of the multipole expansion error in Equation \ref{eq:gerr} is inaccurate for that test. For practical numbers of multipoles \citep[$N_g\geq 6$ - for example, $N_g=8$ in the ABACUS code][]{abacus} the Green's function error is below $L_{\rm MAX}=10^{-3}$ or $L_2=10^{-4}$, which is significantly smaller than the similar error in the Tree-PM approach \citep{gadget4}. 

\begin{figure}[t]
\includegraphics[width=\hsize]{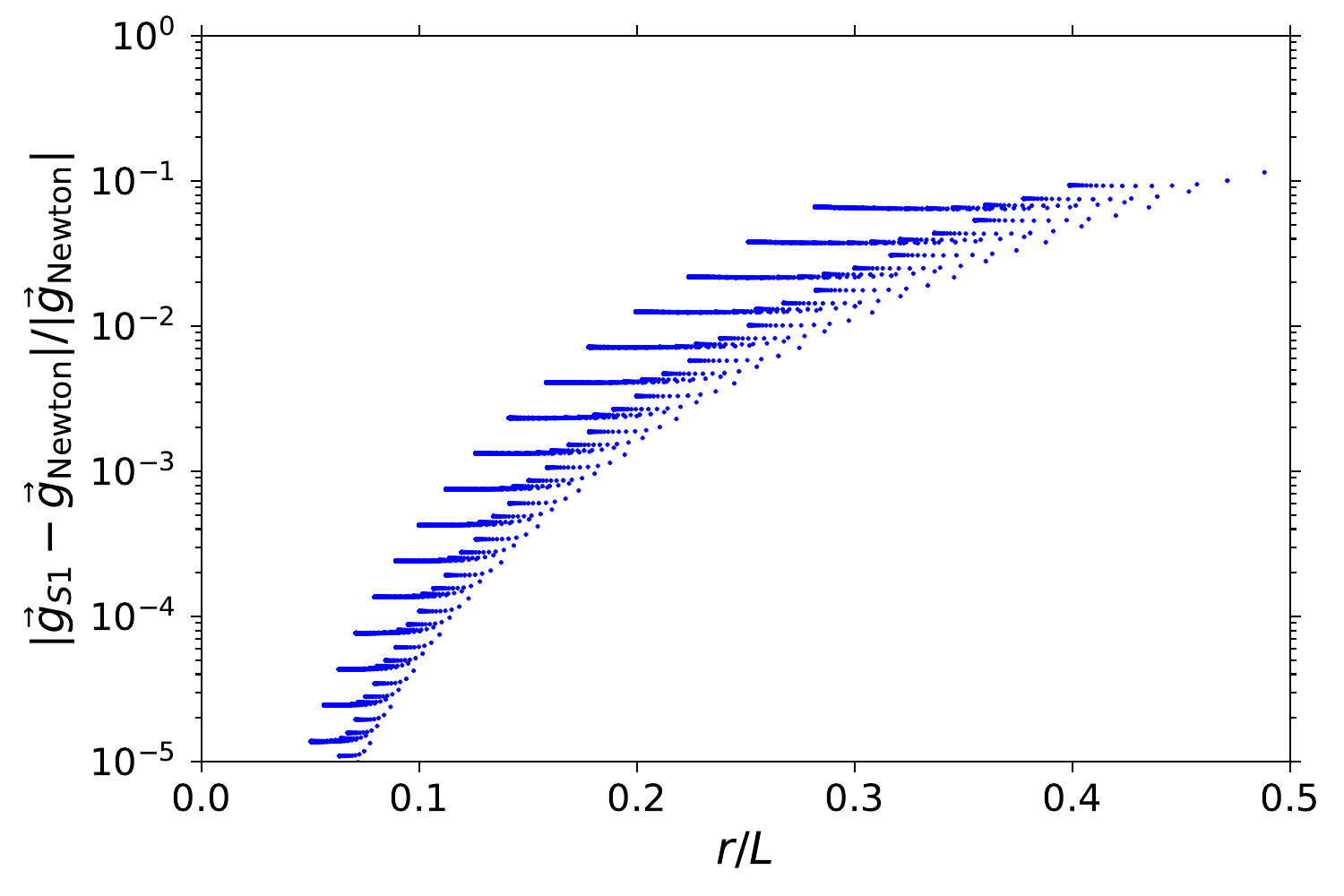}
\caption{Deviation of the gravitational acceleration from the Newtonian form for the approximately optimal Green's function $G_{S1}$. This plot can be compared directly with Fig.\ 2 from GADGET-4 paper \citep{gadget4}.\label{fig:rgs}} 
\end{figure}

Figure \ref{fig:rgs} shows the deviations of the FMM gravitational acceleration computed using the $G_{S1}$ Green's function from the exactly Newtonian form. A similar comparison for the Tree-PM method is presented in \citet{gadget4}. Deviations from the Newtonian form are not perfectly radial as in Tree-PM, but generally significantly smaller.

\begin{deluxetable}{llll}[t]
\tablecaption{Best-fit parameters for $G_S$\label{tab:ak}}
\tablehead{\colhead{$N_g$} & \colhead{$L_{\rm MAX}$} & \colhead{$L_2$} & \colhead{$a_k$}}
\startdata
\cutinhead{2D}
4 & 0.00216 & 0.000514 & -0.124297 \\
6 & 6.73e-05 & 1.36e-05 & -0.120703 \\
8 & 1.84e-05 & 2.52e-06 & -0.120703 \\
\cutinhead{3D}
4 & 0.0112 & 0.00126 & -0.138598 \\
6 & 0.000648 & 6.37e-05 & -0.120236 \\
8 & 0.000192 & 6.00e-06 & -0.120236 \\
\enddata
\end{deluxetable}

\section{Conclusions}

The Green's function
\begin{equation}
    G_{S1}(\vec{x}) \equiv -\frac{1}{4\pi} g\left(\left|\vec{\hat{x}}\diamond L\right|\right)
    \label{eq:gfs1}
\end{equation}
with
\[
  \hat{x} = \frac{9L}{8\pi}\sin(\pi x/L) - \frac{L}{24\pi}\sin(3\pi x/L)
\]
is an approximately optimal Green's function for an implementation of the FMM algortithm that supports periodic boundary conditions "natively", without computing lattice sums or relying on hybrid approaches like FMM-PM.

Where one would go from here? For many practical applications $G_{S1}$ would be enough as it is already more than 100 times more accurate in $L_2$ sense and 20 times more accurate in $L_{]rm MAX}$ sense than the GADGET-4 Tree-PM or FMM-PM approach \citep[][, Fig. 18]{gadget4}. However, its primary limitation is that there is no clear path to improving its precision. I have also tried several other functional forms for $G_*$, including Taylor expansion of $G_*$ in powers of $\sin(fx)$ with $f$ as another parameter, but found none that improves upon $G_{S1}$ in any significant way. It does not mean that the improvement is not possible, of course, just that the correct functional form for $G_*$ has not been found yet.

One can imagine alternative approaches to higher precision. All minimization done in this paper relies on standard local minimum finders. Method for searching beyond the local minimum, such as Simulated Annealing, may produce better results. Another possible approach is to parameterize $G_*$ with a very large number of parameters - for example, as a grid of values with every value being its own parameter. Minimizing a such extremely large dimensional problem is difficult, but is likely to lead to much higher precision. For example, in the FMM-HPM algorithm the Green's function only needs to be sampled on the $(4N_g)^D$ grid, and for a realistic number of multipoles $N_g=8$ in 3D this is $32^3=32{,}768$ values, a large minimization problem but certainly not beyond the modern capabilities. With the grid model for $G_*$ it should be possible to find $G_*$ exactly, i.e.\ with zero error. 

As a final note, Fig.\ \ref{fig:eng} has an interesting feature - the error due to periodic Green's function continues to fall as the number of multipoles increases. Since $G_{S1}$ is approximate, one would expect the error to saturate for sufficiently large $N_g$, reflecting the imperfect accuracy of $G_{S1}$. This is not the case, however. One can hypothesize that, perhaps, $G_{S1}$ is then the true limit of $G_*$ for $N_g\rightarrow\infty$. Unfortunately, it is not possible to test such a hypothesis without a method for solving equation (\ref{eq:opt}).

\acknowledgments
I am grateful to Andrey Kravtsov for useful suggestions during the work on this project and to the anonymous referee for constructive comments that significantly improved the original manuscript.
This manuscript has been authored by Fermi Research Alliance, LLC under Contract No. DE-AC02-07CH11359 with the U.S. Department of Energy, Office of Science, Office of High Energy Physics. 

\bibliographystyle{aasjournal}
\bibliography{main}

\end{document}